\documentclass[aps,pre,twocolumn,superscriptaddress]{revtex4}
\usepackage{amsmath, amssymb}
\usepackage{graphicx}
\usepackage{latexsym}

\begin{document}

\title{Survival probability and first-passage-time statistics of a Wiener process\\
driven by an exponential time-dependent drift}

\author{Eugenio Urdapilleta}
\email[]{urdapile@ib.cnea.gov.ar} \affiliation{Divisi\'on de
F\'isica Estad\'istica e Interdisciplinaria \& Instituto Balseiro,
Centro At\'omico Bariloche, Av. E. Bustillo Km 9.500, S. C. de
Bariloche 8400, R\'io Negro, Argentina}

\begin{abstract}
The survival probability and the first-passage-time statistics are
important quantities in different fields. The Wiener process is
the simplest stochastic process with continuous variables, and
important results can be explicitly found from it. The presence of
a constant drift does not modify its simplicity; however, when the
process has a time-dependent component the analysis becomes
difficult. In this work we analyze the statistical properties of
the Wiener process with an absorbing boundary, under the effect of
an exponential time-dependent drift. Based on the backward
Fokker-Planck formalism we set the time-inhomogeneous equation and
conditions that rule the diffusion of the corresponding survival
probability. We propose as the solution an expansion series in
terms of the intensity of the exponential drift, resulting in a
set of recurrence equations. We explicitly solve the expansion up
to second order and comment on higher-order solutions. The
first-passage-time density function arises naturally from the
survival probability and preserves the proposed expansion.
Explicit results, related properties, and limit behaviors are
analyzed and extensively compared to numerical simulations.
\end{abstract}

\maketitle

\section{Introduction}
\indent Since the primeval discovery of the Brownian motion and
its statistical analysis, the variety of applications in which a
relevant stochastic description result is obtained is steadily
increasing \cite{vankampen,hanggi2005}. The Wiener process and the
Ornstein-Uhlenbeck process are idealized statistical descriptions
that apply to many systems. One of the most valuable theoretical
tools to deal with these and other stochastic processes is the
Fokker-Planck (FP) formalism \cite{risken,gardiner}. In this
framework different realizations of a system are described by the
probability density to find the system in a given state at a
certain time, and a diffusion equation describes its temporal
evolution. Two related questions of wide interest in several areas
are the probability that the system remains in a certain domain at
a given time and the instant at which the system leaves it for the
first time. Given the stochastic nature of the process, different
realizations of the system leave this survival domain at different
times and it is natural to consider what the statistical
properties of this random variable are. This question constitutes
the so-called first-passage-time (FPT) problem
\cite{vankampen, risken, gardiner, ricciardi, siegert1951}.\\
\indent The survival probability as well as the FPT problem is
easy to formulate but difficult to tackle, except for some simple
cases. In particular, for a Wiener process driven by a constant
and positive drift toward a fixed positive boundary, these
quantities have a simple analytical solution \cite{risken,
gardiner, ricciardi, siegert1951, tuckwell, gerstein1964}.
However, the extension to a time-dependent drift is not
straightforward, mainly because the system is no longer
time-homogeneous. In one-dimensional systems, the main work on
this topic possibly is \cite{lindner2004}. In that study, the
author extended, via the forward FP description, the classical
results of Siegert \cite{siegert1951} for a particle being driven
by a \textit{small} time-dependent potential, superimposed on a
general field. By applying a perturbation scheme, the author
derives the recurrence relations between the linear corrections of
the moments of the FPT density function. Other series of works
have analyzed the behavior of the system in a time-dependent
sinusoidal drift, in general, studied in the context of stochastic
resonance (see \cite{sr1,sr2,sr3,sr4} for seminal works on this
topic for the Wiener process with an absorbing boundary; for other
processes, we refer the reader to \cite{sr5}). However, we are
interested in the FPT problem of the Wiener process driven by an
exponential time-dependent drift because it naturally arises in
neuroscience, when modeling spiking neurons with adaptation
currents \cite{benda2010}. This process can also be used to model
a neuron with an exponential time-dependent threshold (see
\cite{lindner2005} for the transformation between an exponential
time-dependent drift to an exponential time-dependent threshold).
With reference to moving thresholds, the main related work is
\cite{tuckwell1984}, where the authors analyzed the moments of the
FPT density function of a Markov process with a moving barrier,
giving some specific examples applied to biological sciences.\\
\indent In this work, we study the survival probability and the
FPT density function of the described system in the framework of
the backward FP formalism. We describe the complete statistics,
instead of focusing on its moments as in previous studies. We
obtain the equation and conditions governing the survival
probability and propose a solution in terms of an expansion in the
exponential drift intensity. This results in an infinite set of
recurrence equations, which we explicitly solve up to second
order. Higher-order terms are outlined and discussed. In
particular, we show that all order functions exist and depend
exclusively on the actual time difference when the initial
conditions are imposed for the backward state, as physical
considerations require. This constitutes the exact solution of the
problem. From the knowledge of the survival probability it is
straightforward to derive the \textit{complete} FPT density
function, which in turn results in an expansion series. Since it
is natural to solve the equations via a Laplace transformation, we
review some important related properties easy to compute from the
Laplace transform of the FPT density function.\\
\indent In the second part of this work, we focus on the explicit
results we have obtained and compare them with numerical
simulations. Given that truncation of the expansion series results
in an approximate solution, we mostly analyze the system in the
linear regime. Related properties and the behavior of the linear
solution in different limits are also considered.

\section{Theory}\label{theory}
\indent In one dimension, the FPT problem can be basically
formulated as follows: a state variable evolves stochastically
according to a given law in its phase space, and we are interested
in describing when this variable leaves a certain domain for the
first time. To deal with this problem a number of different
methods or approaches had been described, mostly based on the
knowledge of the time-dependent probability density or its
temporal evolution \cite{risken, gardiner, ricciardi, siegert1951,
tuckwell}. Here, we first solve the survival probability in terms
of the backward FP equation and then derive the FPT density.

\subsection{Survival probability}
\indent The nonautonomous system we address here can be described
in terms of the Langevin equation,

\begin{equation}\label{eq1}
   \frac{dx}{dt} = \mu + \frac{\epsilon}{\tau_{\text{d}}}
   ~\text{e}^{-(t-t_0)/\tau_{\text{d}}} + \xi(t),
\end{equation}

\noindent where $x$ is the state variable (position, voltage,
etc.; hereafter, the position), $t$ is the time, $\mu$ is the
constant part of the drift, $\epsilon$ quantifies the strength of
the exponential time-dependent drift with time constant
$\tau_{\text{d}}$, and $\xi(t)$ is a Gaussian white noise
characterized by $\langle \xi(t) \rangle = 0$ and $\langle \xi(t)
\xi(t') \rangle = 2D~\delta(t-t')$, with $D$ as a constant.\\
\indent Suppose we have a particle at position $x_0$ at time $t_0$
and it evolves to a position $x'$ at a posterior time $t'$
($t'>t_0$) according to a transition probability density
$P(x',t'|x_0,t_0)$. Clearly, in a FPT problem a certain region of
the domain is forbidden and actually the transition probability
density has implicitly incorporated this fact. In this work, we
analyze the region defined by a constant boundary, $x' <
x_{\text{thr}}$, which set the survival domain,
so the forbidden region is $x \geq x_{\text{thr}}$.\\
\indent The survival probability $F(t'|x_0,t_0)$ is the
probability that the particle remains in the survival domain at
time $t'$ given the initial conditions, and it is given simply by
integration of the transition probability in the $x'$ domain

\begin{equation}\label{eq2}
   F(t'|x_0,t_0) = \int_{-\infty}^{x_{\text{thr}}}
   P(x',t'|x_0,t_0)~dx'.
\end{equation}

\indent To describe the transition probability density we use the
backward FP equation. In this case, given that the state variable
is at position $x'$ at time $t'$, the probability density of the
particle being at the position $x$ at an earlier time $t$ ($t<t'$)
is given by

\begin{eqnarray}\label{eq3}
   && \frac{\partial P(x',t'|x,t)}{\partial t} =\nonumber\\
   && \hspace{1.0cm} -\left[\mu+\frac{\epsilon}{\tau_{\text{d}}}
   ~\text{e}^{-(t-t_0)/\tau_{\text{d}}} \right] \frac{\partial
   P(x',t'|x,t)}{\partial x}\nonumber\\
   && \hspace{1.0cm} - D \frac{\partial^2 P(x',t'|x,t)}{\partial
   x^2}.
\end{eqnarray}

\indent The drift coefficient quantifies the first moment of the
differential transition density in the neighborhood of the
backward state $(x,t)$. Necessarily, the local level of the
exponential term is relative to the initial time $t_0$, breaking
up time homogeneity.\\
\indent The FPT is incorporated with the initial and boundary
conditions: $P(x',t'|x,t = t') = 1$ for $x < x_{\text{thr}}$ and
$0$ for $x \geq x_{\text{thr}}$, and
$P(x',t'|x=x_{\text{thr}},t) = 0$.\\
\indent Integration of Eq.~(\ref{eq3}) in $x'$ from $-\infty$ to
$x_{\text{thr}}$ yields the survival probability from time $t$ to
time $t'$:

\begin{eqnarray}\label{eq4}
   \frac{\partial F(t'|x,t)}{\partial t} =
   &-&\left[ \mu+\frac{\epsilon}{\tau_{\text{d}}} ~\text{e}^{-(t-t_0)/\tau_{\text{d}}} \right] \frac{\partial
   F(t'|x,t)}{\partial x} \nonumber \\
   &-& D \frac{\partial^2 F(t'|x,t)}{\partial
   x^2}.
\end{eqnarray}

\indent Since $t'$ is a parameter, we make the substitution $\tau
= t'-t$ and rename the probability $F(x,\tau;t')$. The
corresponding equation is

\begin{eqnarray}\label{eq5}
   && \frac{\partial F(x,\tau;t')}{\partial \tau} = \nonumber \\
   && \hspace{0.5cm} \left[ \mu+\frac{\epsilon}{\tau_{\text{d}}}
   ~\text{e}^{-(t'-t_0)/\tau_{\text{d}}}~\text{e}^{\tau/\tau_{\text{d}}}
   \right]
   \frac{\partial F(x,\tau;t')}{\partial x} \nonumber \\
   && \hspace{0.5cm} + D \frac{\partial^2 F(x,\tau;t')}{\partial
   x^2},
\end{eqnarray}

\noindent with $F(x,\tau=0;t')=1$ for $x<x_{\text{thr}}$ and $0$
for $x\geq x_{\text{thr}}$, and $F(x=x_{\text{thr}},\tau;t')=0$.\\
\indent To solve this equation we propose an expansion in powers
of $\epsilon$:

\begin{eqnarray}\label{eq6}
   F(x,\tau;t') &=& F_{0}(x,\tau;t') + \epsilon F_{1}(x,\tau;t') +
   \epsilon^2 F_{2}(x,\tau;t') + \dots \nonumber\\
   &=& \sum_{n=0}^{\infty} \epsilon^n F_{n}(x,\tau;t').
\end{eqnarray}

\indent Replacing Eq.~(\ref{eq6}) into Eq.~(\ref{eq5}) and
grouping in orders of $\epsilon$, we obtain

\begin{eqnarray}\label{eq7}
   \Bigg[ \frac{\partial F_{0}}{\partial \tau} - \mu \frac{\partial F_0}{\partial
   x} - D \frac{\partial^2 F_{0}}{\partial x^2} \Bigg] \nonumber \hspace{2.1cm}\\
   + ~ \sum_{n=1}^{\infty} \epsilon^{n} ~ \Bigg[ \frac{\partial F_{n}}{\partial \tau} - \mu \frac{\partial F_{n}}{\partial x} - \frac{1}{\tau_{\text{d}}}
   ~\text{e}^{-(t'-t_0)/\tau_{\text{d}}}~\text{e}^{\tau/\tau_{\text{d}}}~
   \frac{\partial F_{n-1}}{\partial x} \nonumber\\
   - D \frac{\partial^2 F_{n}}{\partial
   x^2} \Bigg] = 0, \hspace{1.35cm}
\end{eqnarray}

\noindent where we have simplified the notation for the sake of
clarity.\\
\indent Since $\epsilon$ is a parameter, each term in brackets
should be identically $0$. Therefore, to find the survival
probability we have to solve

\begin{eqnarray}\label{eq8}
   \frac{\partial F_{0}}{\partial \tau} - \mu \frac{\partial F_{0}}{\partial
   x} - D \frac{\partial^2 F_{0}}{\partial x^2} = 0 \hspace{3.82cm}\\
   \label{eq9}
   \frac{\partial F_{n}}{\partial \tau} - \mu \frac{\partial F_{n}}{\partial
   x} - D \frac{\partial^2 F_{n}}{\partial x^2} = \nonumber\hspace{4.0cm}\\
   \frac{1}{\tau_{\text{d}}} ~\text{e}^{-(t'-t_0)/\tau_{\text{d}}}~\text{e}^{\tau/\tau_{\text{d}}}~
   \frac{\partial F_{n-1}}{\partial x} ~~\text{for}~n\geq
   1.\hspace{0.5cm}
\end{eqnarray}

\indent This system of Eqs.~(\ref{eq8}) and (\ref{eq9}), can be
solved recursively up to the degree of accuracy needed. To
complete the solution of the survival probability we have to
define the initial and boundary conditions for all the functions
$F_{n}(x,\tau;t')$, for $n \geq 0$. Again, given the arbitrariness
of $\epsilon$, the nonhomogeneous conditions should be imposed to
the zeroth-order function. Therefore, initial conditions are

\begin{eqnarray}\label{eq10}
   F_{0}(x,\tau=0;t') &=& \left\{ \begin{split} 1 ~~\text{if}~
   x<x_{\text{thr}},\\
   0 ~~\text{if}~x\geq x_{\text{thr}},
   \end{split}\right.\\
   \label{eq11}
   F_{n}(x,\tau=0;t') &=& 0 ~~\text{for}~ n\geq 1,
\end{eqnarray}

\noindent whereas boundary condition is
$F_{n}(x=x_{\text{thr}},\tau;t') = 0$ for $n\geq 0$.\\
\indent Next we solve the expansion up to the second-order term
and analyze higher orders.

\subsubsection{Zeroth order solution} \indent The system described
by Eqs.~(\ref{eq8}) and (\ref{eq10}) corresponds to the constant
drift case ($\epsilon = 0$). The survival probability of the
Wiener process with constant drift and diffusion coefficients is a
time-homogeneous process (the system remains unchanged with a
shift in $t'$) and easy to solve in Laplace domain. Omitting the
dependence in $s$ (to solve the equation, $s$ acts as a
parameter), this probability reads

\begin{equation}\label{eq12}
   \tilde{F}_{0}^{L}(x) = \frac{1}{s} - \frac{1}{s} \exp\Big\{ \frac{(x_{\text{thr}}-x)}{2D}
   \left[\mu - \sqrt{\mu^2 + 4Ds} \right] \Big\},
\end{equation}

\noindent where we denote $\tilde{F}_{0}^{L}(x)$ the Laplace
transform of $F_{0}(x,\tau)$ to the $s$ domain (due to time
homogeneity, $t'$ only appears in $\tau$). In deriving
Eq.~(\ref{eq12}) we have used the fact that $\tilde{F}_{0}^{L}(x
\rightarrow -\infty)$ is bounded.\\
\indent By the inverse Laplace transformation of Eq.~(\ref{eq12}),
we obtain the solution in terms of $\tau = t'-t$, $F_{0}(x,t'-t)$.
At this point we state the initial conditions of the problem, $x =
x_0$ and $t = t_0$. Therefore, $F(x,t'-t) \rightarrow
F(x_0,t'-t_0)$. Again, replacing $\tau = t'-t_0$ (now, $\tau$ is
the actual time difference) and transforming back to the $s$
domain, we obtain

\begin{equation}\label{eq13}
   \tilde{F}_{0}^{L}(s) = \frac{1}{s} - \frac{1}{s} \exp\Big\{ \frac{(x_{\text{thr}}-x_0)}{2D}
   \left[\mu - \sqrt{\mu^2 + 4Ds} \right] \Big\},
\end{equation}

\noindent where now we have recovered the dependence on $s$ in the
notation. By comparing Eqs.~(\ref{eq12}) and (\ref{eq13}) we note
just a single change, $x \rightarrow x_0$. However, the procedure
described is important in time-inhomogeneous problems and it will
be important when solving the following orders.\\
\indent Even when the inverse Laplace transform of
Eq.~(\ref{eq13}) is available, we disregard this step since, as we
will see later when deriving the FPT density function, it is
unnecessary (and actually it is related to the FPT cumulative
distribution).

\subsubsection{First order solution}
\indent The first order term is given by the solution of
Eq.~(\ref{eq9}) for $n=1$,

\begin{eqnarray}\label{eq14}
   \frac{\partial F_{1}(x,\tau;t')}{\partial \tau} - \mu \frac{\partial
   F_{1}(x,\tau;t')}{\partial x} - D \frac{\partial^2 F_{1}(x,\tau;t')}{\partial x^2} =
   \nonumber \\
   \frac{1}{\tau_{\text{d}}}~\text{e}^{-(t'-t_0)/\tau_{\text{d}}}~
   \frac{\partial}{\partial x}
   \left[\text{e}^{\tau/\tau_{\text{d}}}~F_{0}(x,\tau;t')\right],
\end{eqnarray}

\noindent with the corresponding initial and boundary conditions.\\
\indent This equation can be solved via the Laplace transform. In
the $s$ domain, Eq.~(\ref{eq14}) reads

\begin{eqnarray}\label{eq15}
   s ~\tilde{F}_{1}^{L}(x;t') - \mu \frac{d \tilde{F}_{1}^{L}(x;t')}{dx}
   - D \frac{d^2 \tilde{F}_{1}^{L}(x;t')}{dx^2} = \nonumber \hspace{1.0cm}\\
   \frac{1}{\tau_{\text{d}}}~\text{e}^{-(t'-t_0)/\tau_{\text{d}}}
   ~\frac{d}{dx} \big\{ \mathcal{L} \left[F_{0}(x,\tau;t')\right]_{(s)} \big\}_{\rfloor
   s-1/\tau_{\text{d}}},
\end{eqnarray}

\noindent where $\mathcal{L} \left[ \cdot \right]_{(s)}$
represents the Laplace transform operator and
$\tilde{F}_{1}^{L}(x;t')$ is the Laplace transform of
$F_{1}(x,\tau;t')$. Substituting the result we obtained before,
Eq.~(\ref{eq12}) (note that the initial state of the problem is
\textit{not} already evaluated), into Eq.~(\ref{eq15}) we have

\begin{eqnarray}\label{eq16}
   s ~\tilde{F}_{1}^{L}(x;t') - \mu \frac{d \tilde{F}_{1}^{L}(x;t')}{dx}
   - D \frac{d^2 \tilde{F}_{1}^{L}(x;t')}{dx^2} = \nonumber \hspace{1.0cm}\\
   \frac{1}{\tau_{\text{d}}}~\text{e}^{-(t'-t_0)/\tau_{\text{d}}}~\frac{\left[\mu-
   \sqrt{\mu^2+4D(s-1/\tau_{\text{d}})}\right]}{2D(s-1/\tau_{\text{d}})}\hspace{1cm}\nonumber\\
   \cdot~\exp\Big\{\frac{(x_{\text{thr}}-x)}{2D}\left[\mu-\sqrt{\mu^2+4D
   (s-1/\tau_{\text{d}})}\right] \Big\}.
\end{eqnarray}

\indent The general solution to this equation is given by

\begin{eqnarray}\label{eq17}
   \tilde{F}_{1}^{L}(x;t') = \text{C}_{1}~\exp \left( - \frac{\mu+\sqrt{\mu^2 + 4Ds}}{2D}~x \right) \hspace{1.0cm}\nonumber\\
   + ~\text{C}_{2}~\exp \left( - \frac{\mu-\sqrt{\mu^2 + 4Ds}}{2D}~x
   \right)\nonumber\hspace{2.3cm}\\
   + ~\frac{1}{2D}~\text{e}^{-(t'-t_0)/\tau_{\text{d}}}
   ~\frac{\left[\mu-\sqrt{\mu^2+4D(s-1/\tau_{\text{d}})}\right]}{(s-1/\tau_{\text{d}})}\hspace{0.48cm}\nonumber\\
   \cdot~\exp\Big\{\frac{(x_{\text{thr}}-x)}{2D}\left[\mu-\sqrt{\mu^2+4D
   (s-1/\tau_{\text{d}})}\right] \Big\},
\end{eqnarray}

\noindent valid for $\text{Re}(s) \geq 1/\tau_{\text{d}}$.\\
\indent Taking into account that $\tilde{F}_{1}^{L}(x\rightarrow
-\infty;t')$ is bounded and the boundary condition is
$\tilde{F}_{1}^{L}(x=x_{\text{thr}};t') = 0$, we obtain

\begin{eqnarray}\label{eq18}
   \tilde{F}_{1}^{L}(x;t') =
   \frac{1}{2D}~\text{e}^{-(t'-t_0)/\tau_{\text{d}}}~\frac{\left[\mu-\sqrt{\mu^2+4D(s-1/\tau_{\text{d}})}\right]}{(s-1/\tau_{\text{d}})}\nonumber\\
   \cdot~\Bigg\{ ~~\exp\Big\{ \frac{(x_{\text{thr}}-x)}{2D}\left[ \mu-\sqrt{\mu^2+4D(s-1/\tau_{\text{d}})} \right]\Big\} \nonumber\\
   -~\exp\Big\{ \frac{(x_{\text{thr}}-x)}{2D}\left[ \mu-\sqrt{\mu^2+4Ds} \right] \Big\}
   \Bigg\}.\hspace{1.0cm}
\end{eqnarray}

\indent We further operate with the inverse Laplace transform of
Eq.~(\ref{eq18}), which is

\begin{eqnarray}\label{eq19}
   F_{1}(x,\tau;t') = \frac{1}{2D}~\text{e}^{-(t'-t_0)/\tau_{\text{d}}}\hspace{3.5cm}\nonumber\\
   ~\cdot\frac{1}{2\pi j}
   \int_{\sigma-j\infty}^{\sigma+j\infty}~\text{e}^{s\tau}~
   \frac{\left[\mu-\sqrt{\mu^2+4D(s-1/\tau_{\text{d}})}\right]}{(s-1/\tau_{\text{d}})}\hspace{0.671cm}\nonumber\\
   \cdot~\Bigg\{ ~~\exp\Big\{ \frac{(x_{\text{thr}}-x)}{2D}\left[ \mu-\sqrt{\mu^2+4D(s-1/\tau_{\text{d}})} \right]\Big\} \nonumber\\
   -~\exp\Big\{ \frac{(x_{\text{thr}}-x)}{2D}\left[ \mu-\sqrt{\mu^2+4Ds} \right] \Big\}
   \Bigg\}~ds,\hspace{0.5cm}
\end{eqnarray}

\noindent where $j$ represents the imaginary unit and $\sigma \geq
1/\tau_{\text{d}}$. Taking the substitution $z =
s-1/\tau_{\text{d}}$, we obtain

\begin{eqnarray}\label{eq20}
   F_{1}(x,\tau;t') = \frac{1}{2D}~\text{e}^{-(t'-t_0)/\tau_{\text{d}}}~\text{e}^{~\tau/\tau_{\text{d}}}\hspace{2.5cm}\nonumber\\
   ~\cdot\frac{1}{2\pi j}
   \int_{\sigma_{z}-j\infty}^{\sigma_{z}+j\infty}~\text{e}^{z \tau}~
   \frac{\left[\sqrt{\mu^2+4Dz}-\mu \right]}{z}\hspace{1.9cm}\nonumber\\
   \cdot~\Bigg\{ ~~\exp\Big\{ \frac{(x_{\text{thr}}-x)}{2D}\left[ \mu-\sqrt{\mu^2+4D(z+1/\tau_{\text{d}})} \right]\Big\} \nonumber\\
   -~\exp\Big\{ \frac{(x_{\text{thr}}-x)}{2D}\left[ \mu-\sqrt{\mu^2+4Dz} \right] \Big\}
   \Bigg\}~dz,\hspace{0.52cm}
\end{eqnarray}

\noindent where now, it is easy to check that the region of
convergence of the integrand is $\text{Re}(z) = \sigma_{z} \geq
0$. However, there still is an exponential factor that makes the
expression diverge.\\
\indent At this point we are able to evaluate the real conditions
of the problem: $x = x_0$ and $t = t_0$. Remembering that $\tau =
t'-t$, the latter condition imposes that the two exponential
factors (before the integral) in Eq.~(\ref{eq20}) cancel each
other. Hereafter, we use $\tau$ to represent the actual time
referred to the initial time, $\tau = t'-t_0$. Taking the Laplace
transform on this variable, from Eq.~(\ref{eq20}) the function
$F_{1}(\tau)$ (note that $x$ was evaluated and the dependence on
$t'$ is exclusively given by the combination in $\tau$) transforms
to

\begin{eqnarray}\label{eq21}
   \tilde{F}_{1}^{L}(s) = \frac{1}{2D}
   \frac{\left[\sqrt{\mu^2+4Ds}-\mu \right]}{s}\hspace{3.2cm}\nonumber\\
   \cdot~\Bigg\{ ~~\exp\Big\{ \frac{(x_{\text{thr}}-x_0)}{2D}\left[ \mu-\sqrt{\mu^2+4D(s+1/\tau_{\text{d}})} \right]\Big\} \nonumber\\
   -~\exp\Big\{ \frac{(x_{\text{thr}}-x_0)}{2D}\left[ \mu-\sqrt{\mu^2+4Ds} \right] \Big\}
   \Bigg\},\hspace{1.0cm}
\end{eqnarray}

\noindent valid for $\text{Re}(s) \geq 0$.\\

\subsubsection{A note on the higher-order solutions}
\indent In this subsection we remark on some aspects of the
existence and the convergence of higher-order terms expressed in
the Laplace domain. The higher-order terms in the expansion,
Eq.~(\ref{eq6}), correspond to the solution of Eq.~(\ref{eq9})
with the appropriate initial and boundary conditions ($n \geq 2$).
In particular, we obtain an equation analogous to
Eq.~(\ref{eq14}):

\begin{eqnarray}\label{eq22}
   \frac{\partial F_{n}(x,\tau;t')}{\partial \tau} - \mu \frac{\partial
   F_{n}(x,\tau;t')}{\partial x} - D \frac{\partial^2 F_{n}(x,\tau;t')}{\partial x^2} =
   \nonumber \\
   \frac{1}{\tau_{\text{d}}}~\text{e}^{-(t'-t_0)/\tau_{\text{d}}}~
   \frac{\partial}{\partial x}
   \left[\text{e}^{\tau/\tau_{\text{d}}}~F_{n-1}(x,\tau;t')\right].
\end{eqnarray}

\indent The term on the right-hand side of the equation
corresponds to a source because $F_{n-1}(x,\tau;t')$ was already
solved. As in the first-order case, the knowledge of the source
term in the $s$ domain enables us to readily Laplace transform the
equation, obtaining an ordinary differential equation with a
forcing term. The homogeneous part of the solution is exactly the
same as that in Eq.~(\ref{eq17}) (terms with unknown constants
$C_{i}$) and the particular solution is different for different
orders. Moreover, $C_{1}$ has to be $0$ for bounded solutions and
the existence of the particular solution is given as a sum of the
exponential factors present in $\tilde{F}_{n-1}^{L}(x;t')
\rfloor_{s-1/\tau_{\text{d}}}$. After evaluation of the boundary
condition, the solution is given as the sum of
$n+1$ exponential terms.\\
\indent Here we note the structure that this forcing term imposes
on the solution. Since Eq.~(\ref{eq22}) operates in the backward
state $(x,t)$, the previous-order solution $F_{n-1}(x,\tau;t')$
must not be evaluated in the initial state $(x_0,t_0)$, or
correspondingly, its Laplace transform should be
done for the variable $\tau = t'-t$ and not for $\tau = t'-t_0$.\\
\indent To simplify, we exemplify the concepts of convergence with
the second-order solution and then extend the conclusion to all
orders. In the Laplace domain, the equation governing the
second-order solution is

\begin{eqnarray}\label{ad1}
   s ~\tilde{F}_{2}^{L}(x;t') - \mu \frac{d \tilde{F}_{2}^{L}(x;t')}{dx}
   - D \frac{d^2 \tilde{F}_{2}^{L}(x;t')}{dx^2} = \nonumber \hspace{1.0cm}\\
   \frac{1}{\tau_{\text{d}}}~\text{e}^{-(t'-t_0)/\tau_{\text{d}}}
   ~\frac{d}{dx} \big\{ \mathcal{L} \left[F_{1}(x,\tau;t')\right]_{(s)} \big\}_{\rfloor
   s-1/\tau_{\text{d}}},
\end{eqnarray}

\noindent where the Laplace transform of the previous-order
solution is given by Eq.~(\ref{eq18}). Prior to the evaluation of
the initial state, this solution has a region of convergence
$\text{Re}(s) \geq 1/\tau_{\text{d}}$. It is easy to check that,
due to the delay introduced in the Laplace domain, the forcing
term in Eq.~(\ref{ad1}) will impose that the region of convergence
of the Laplace transform of the second-order solution is
$\text{Re}(s) \geq 2/\tau_\text{d}$, but now a factor
$\exp[-2(t'-t_0)/\tau_{\text{d}}]$ appears. Explicitly, the
equation governing the second-order is given by

\begin{eqnarray}\label{ad2}
   s ~\tilde{F}_{2}^{L}(x;t') - \mu \frac{d \tilde{F}_{2}^{L}(x;t')}{dx}
   - D \frac{d^2 \tilde{F}_{2}^{L}(x;t')}{dx^2} = \hspace{1.0cm}\nonumber\\
   \frac{1}{2D\tau_{\text{d}}}~\text{e}^{-2(t'-t_0)/\tau_{\text{d}}}~\frac{\left[\mu-
   \sqrt{\mu^2+4D(s-2/\tau_{\text{d}})}\right]}{2D(s-2/\tau_{\text{d}})}
   \hspace{0.4cm}\nonumber\\
   \cdot~\Bigg\{ \left[\mu-\sqrt{\mu^2+4D(s-1/\tau_{\text{d}})}\right]
   \hspace{3.0cm}\nonumber\\
   \cdot \exp\Big\{\frac{(x_{\text{thr}}-x)}{2D}\left[\mu-\sqrt{\mu^2+4D
   (s-1/\tau_{\text{d}})}\right] \Big\} \hspace{0.38cm}\nonumber\\
   -\left[\mu-\sqrt{\mu^2+4D(s-2/\tau_{\text{d}})}\right]
   \hspace{3.0cm}\nonumber\\
   \cdot \exp\Big\{\frac{(x_{\text{thr}}-x)}{2D}\left[\mu-\sqrt{\mu^2+4D
   (s-2/\tau_{\text{d}})}\right] \Big\} \Bigg\},
\end{eqnarray}

\noindent and its solution is

\begin{eqnarray}\label{ad3}
   \tilde{F}_{2}^{L}(x;t') = \frac{1}{2D}~\text{e}^{-2(t'-t_0)/\tau_{\text{d}}}~
   \frac{\left[\mu-\sqrt{\mu^2+4D(s-2/\tau_{\text{d}})}\right]}{2D(s-2/\tau_{\text{d}})}\nonumber\\
   \cdot \sum_{i=0}^{2}
   a_{i}~\exp\Big\{\frac{(x_{\text{thr}}-x)}{2D}\left[\mu-\sqrt{\mu^2+4D(s-i/\tau_{\text{d}})}\right]\Big\},\hspace{0.4cm}
\end{eqnarray}

\noindent where

\begin{eqnarray}\label{ad4}
   a_{0} &=&
   \sqrt{\mu^2+4D(s-1/\tau_{\text{d}})} -
   \frac{1}{2}\sqrt{\mu^2+4D(s-2/\tau_{\text{d}})} \nonumber\\
   & & - \frac{1}{2}\mu, \nonumber\\
   a_{1} &=& \mu - \sqrt{\mu^2+4D(s-1/\tau_{\text{d}})},   \nonumber\\
   a_{2} &=& - \frac{1}{2} \left[ \mu - \sqrt{\mu^2+4D(s-2/\tau_{\text{d}})} \right].
\end{eqnarray}

\indent Proceeding as in Eqs.~(\ref{eq19}) and (\ref{eq20}), we
obtain two exponential factors, $\exp[-2(t'-t_0)/\tau_{\text{d}}]$
and $\exp(2\tau/\tau_{\text{d}})$, that cancel each other when the
initial state $(x_0,t_0)$ is imposed. This cancellation actually
means that the second-order term of the survival probability, with
the initial state imposed, depends on time exclusively through the
combination $\tau = t'-t_0$ (actual time difference). Therefore,
its Laplace transform (on the variable $\tau = t'-t_0$) with the
initial state evaluated is

\begin{eqnarray}\label{ad5}
   \tilde{F}_{2}^{L}(s) = \frac{1}{2D}~\frac{\left[ \mu-\sqrt{\mu^2+4Ds}
   \right]}{2Ds} \sum_{i=0}^{2} b_{i}(s)\hspace{2.0cm}\nonumber\\
   \cdot
   \exp\Big\{\frac{(x_{\text{thr}}-x_0)}{2D}
   \left[\mu-\sqrt{\mu^2+4D(s+i/\tau_{\text{d}})}\right]
   \Big\},\hspace{0.4cm}
\end{eqnarray}

\noindent where

\begin{eqnarray}\label{ad6}
   b_{0}(s) &=& -\frac{1}{2} \left[ \mu - \sqrt{\mu^2+4Ds}
   \right],\nonumber\\
   b_{1}(s) &=&
   \mu-\sqrt{\mu^2+4D(s+1/\tau_{\text{d}})},\nonumber\\
   b_{2}(s) &=& \sqrt{\mu^2+4D(s+1/\tau_{\text{d}})} -
   \frac{1}{2}\sqrt{\mu^2+4Ds} \nonumber\\
   & & - \frac{1}{2}\mu.
\end{eqnarray}

\indent Recursively, the $n$th-order solution in the backward
state $(x,t)$ has a Laplace transform valid for $\text{Re}(s) \geq
n/\tau_{\text{d}}$ with a factor
$\exp[-n(t'-t_0)/\tau_{\text{d}}]$. Therefore, the preceding
conclusion extends to all orders.\\
\indent Given the existence of the solution of all terms, the
expansion proposed in Eq.~(\ref{eq6}) constitutes the exact
solution of the system.

\subsection{First-passage time density}
\indent In the previous subsection, we demonstrate that the
expansion given by Eq.~(\ref{eq6}), with the initial state
evaluated, $(x_0,t_0)$, actually reads

\begin{eqnarray}\label{eq23}
   F(x_0,\tau) &=& F_{0}(x_0,\tau) + \epsilon F_{1}(x_0,\tau) +
   \epsilon^2 F_{2}(x_0,\tau) + \dots \nonumber\\
   &=& \sum_{n=0}^{\infty} \epsilon^n F_{n}(x_0,\tau),
\end{eqnarray}

\noindent where the dependence on time appears exclusively through
the combination $\tau = t'-t_0$.\\
\indent Once the initial conditions are stated, by definition,
$F(x_0,\tau)$ is the probability that the particle remains at time
$\tau = t'-t_0$ in the survival domain and, hence, equals the
probability that the FPT is posterior to $\tau$: $F(x_0,\tau) =
\text{Prob}(T> \tau)$, where $T$ represents the FPT. In terms of
the cumulative distribution function of the FPT random variable,
$\Phi(\tau),$ this means that $F(x_0,\tau) = 1 - \Phi(\tau)$
(hereafter, $x_0$ is a parameter and can be disregarded from
notation). The density function, $\phi(\tau)$, is given by

\begin{equation}\label{eq24}
   \phi(\tau) = \frac{d \Phi(\tau)}{d\tau} = -\frac{\partial F(x_0,\tau)}{\partial
   \tau},
\end{equation}

\noindent which means that the FPT density function has an
expansion given by

\begin{equation}\label{eq25}
   \phi(\tau) = - \sum_{n=0}^{\infty} \epsilon^n ~ \frac{\partial F_{n}(x_0,\tau)}{\partial
   \tau}.
\end{equation}

\indent Remembering that the initial condition in the diffusion
problem reads $F(x,\tau=0;t') = 1$ for $x < x_{\text{thr}}$, it
results that $F(x_0,\tau=0) = 1$ in the solution already evaluated
with the conditions of the problem (and obviously $x_0 <
x_{\text{thr}}$ for a nontrivial problem). Therefore, the Laplace
transform of Eq.~(\ref{eq24}) reads

\begin{equation}\label{eq26}
   \tilde{\phi}^{L}(s) = 1 - s~\tilde{F}^{L}(x_0,s),
\end{equation}

\noindent where $\tilde{\phi}^{L}(s)$ [$\tilde{F}^{L}(x_0,s)$] is
the Laplace transform of $\phi(\tau)$ [$F(x_0,\tau)$].\\
\indent Equivalently, in terms of the expansion for
$\tilde{F}^{L}(x_0,s)$ [see Eq.~(\ref{eq23})], the Laplace
transform of the density is

\begin{equation}\label{eq27}
   \tilde{\phi}^{L}(s) = 1 - s ~ \sum_{n=0}^{\infty} \epsilon^n ~
   \tilde{F}_{n}^{L}(x_0,s),
\end{equation}

\noindent where $\tilde{F}_{n}^{L}(x_0,s)$ is the Laplace
transform of the $n$th term in the expansion of $F(x_0,\tau)$,
$F_{n}(x_0,\tau)$.\\
\indent Since $\phi(\tau)$ has an expansion given by
Eq.~(\ref{eq25}), it is natural to write

\begin{equation}\label{eq28}
   \phi(\tau) = \sum_{n=0}^{\infty} \epsilon^n ~ \phi_{n}(\tau),
\end{equation}

\noindent where

\begin{eqnarray}\label{eq29}
   \phi_{n}(\tau) &=& -\frac{\partial F_{n}(x_0,\tau)}{\partial \tau}.
\end{eqnarray}

\indent In the $s$ domain, Eq.~(\ref{eq28}) reads

\begin{equation}\label{eq30}
   \tilde{\phi}^{L}(s) = \sum_{n=0}^{\infty} \epsilon^n ~
   \tilde{\phi}_{n}^{L}(s),
\end{equation}

\noindent where $\tilde{\phi}_{n}^{L}(s)$ is the Laplace transform
of the $n$th term in the expansion of $\phi(\tau)$,
$\phi_{n}(\tau)$, and it is given by

\begin{eqnarray}\label{eq31}
   \tilde{\phi}_{0}^{L}(s) &=& 1 - s~\tilde{F}_{0}^{L}(x_0,s),\\
   \label{eq32}
   \tilde{\phi}_{n}^{L}(s) &=& -
   s~\tilde{F}_{n}^{L}(x_0,s),~~\text{for}~n\geq 1.
\end{eqnarray}

\indent For example, from the findings in the previous subsection,
the terms in the expansion up to the first order of the FPT
density function are

\begin{eqnarray}\label{eq33}
   \tilde{\phi}_{0}^{L}(s) = \exp\Big\{ \frac{(x_{\text{thr}}-x_0)}{2D}
   \left[\mu - \sqrt{\mu^2 + 4Ds} \right] \Big\},
\end{eqnarray}

\begin{eqnarray}\label{eq34}
   \tilde{\phi}_{1}^{L}(s) = \frac{\left[\mu -
   \sqrt{\mu^2+4Ds}\right]}{2D}\hspace{3.2cm}\nonumber\\
   \cdot~\Bigg\{ ~~\exp\Big\{ \frac{(x_{\text{thr}}-x_0)}{2D}\left[ \mu-\sqrt{\mu^2+4D(s+1/\tau_{\text{d}})} \right]\Big\} \nonumber\\
   -~\exp\Big\{ \frac{(x_{\text{thr}}-x_0)}{2D}\left[ \mu-\sqrt{\mu^2+4Ds} \right] \Big\}
   \Bigg\}.\hspace{1.0cm}
\end{eqnarray}

\indent Eq.~(\ref{eq33}) is the classical result for the FPT
problem with constant drift $\mu$ and diffusion $D$ coefficients
\cite{tuckwell}, consistent with our approach.

\subsection{Related properties of the first-passage time
density}\label{prop} \indent Since the solution of the proposed
expansion is naturally obtained in the Laplace domain, here we
review some properties easy to calculate from this knowledge. It
is easy to check that the moments of the density function satisfy

\begin{equation} \label{eq35}
\langle \tau^k \rangle = \int_0^{\infty} \phi(\tau) ~ \tau^k ~
d\tau = (-1)^k \frac{d^k\tilde{\phi}^{L}(s)}{ds^k}\rfloor_{s=0},
\end{equation}

\noindent which means that all the moments preserve the expansion
in $\epsilon$

\begin{equation} \label{eq36}
   \langle \tau^k \rangle = \sum_{n=0}^{\infty} \epsilon^{n} ~
\langle \tau^k \rangle_{\phi_{n}},
\end{equation}

\noindent where

\begin{equation}\label{eq37}
   \langle \tau^k \rangle_{\phi_n} = (-1)^k
   \frac{d^k\tilde{\phi}_{n}^{L}(s)}{ds^k}\rfloor_{s=0}.
\end{equation}

\indent For example, the first two moments for the unperturbed
case ($n=0$) are

\begin{eqnarray}\label{eq38}
   \langle \tau \rangle_{\phi_0} &=&
   \frac{x_{\text{thr}}-x_0}{\mu}, \nonumber \\
   \langle \tau^2 \rangle_{\phi_0} &=&
   \frac{2D(x_{\text{thr}}-x_0)}{\mu^3} +
   \frac{(x_{\text{thr}}-x_0)^2}{\mu^2},
\end{eqnarray}

\noindent which correspond to the constant drift case
\cite{tuckwell}.\\
\indent The linear changes in these properties, Eq.~(\ref{eq36}),
for $n~=~1$, are

\begin{eqnarray}\label{eq39}
   \langle \tau \rangle_{\phi_1} &=& \frac{1}{\mu} \nonumber\\
   &\cdot& \Bigg\{ \exp\left[
   \frac{(x_{\text{thr}}-x_0)}{2D}\left(\mu-\sqrt{\mu^2+4D/\tau_{\text{d}}}\right)
   \right]-1\Bigg\},\nonumber\\
   \langle \tau^2 \rangle_{\phi_1} &=& \frac{2}{\mu^2} \nonumber\\
   &\cdot& \Bigg\{ \left[
   \frac{\mu(x_{\text{thr}}-x_0)}{\sqrt{\mu^2+4D/\tau_{\text{d}}}}+\frac{D}{\mu}
   \right] \nonumber\\
   && ~~~\cdot\exp\left[
   \frac{(x_{\text{thr}}-x_0)}{2D}\left(\mu-\sqrt{\mu^2+4D/\tau_{\text{d}}}\right)
   \right]\nonumber\\
   && ~~~- (x_{\text{thr}}-x_0) - \frac{D}{\mu} \Bigg\}.
\end{eqnarray}

\indent These results, Eq.~(\ref{eq39}), coincide with those of
the corresponding case in \cite{lindner2004}, obtained from a
different approach.\\
\indent The assessment of the complete density function in the
Laplace domain enables us to obtain another important property.
The successive ordering of FPTs of Wiener processes, each of them
independent of the history (in our system, this means the fixed
escape domain and initial state), constitutes a renewal process.
Given the times $\{t_{k}\}$ when the system reaches the threshold
$x_{\text{thr}}$ starting from $x_0$ and setting it again to
$x_0$, we can construct a ``spike train'', $X(t)$, defined by

\begin{equation}\label{eq40}
   X(t) = \sum_{\{t_{k}\}} \delta(t-t_{k}),
\end{equation}

\noindent representing a renewal point process.\\
\indent The Fourier transform of $X(t)$ is the spike train
spectral density $S(\omega)$, which represents an important
property in some fields, such as neuroscience \cite{stein1972}. It
is related to the density function of a \textit{single} escape
process, expressed in the Laplace domain, through
\cite{cox1965,stein1972}

\begin{equation} \label{eq41}
S(\omega) = \frac{1}{2\pi\langle \tau \rangle} \Big[ 1 +
\frac{\tilde{\phi}^{L}(j\omega)}{1-\tilde{\phi}^{L}(j\omega)} +
\frac{\tilde{\phi}^{L}(-j\omega)}{1-\tilde{\phi}^{L}(-j\omega)}\Big],
\end{equation}

\noindent where $\omega$ is the angular frequency. The sequence of
the renewal times is mean subtracted; otherwise, a $\delta$ peak
appears at frequency $0$.\\
\indent In particular, the constant driving case ($\epsilon = 0$)
is analytically tractable and a relatively simple expression is
found in \cite{stein1972}. The exponential driving case ($\epsilon
\neq 0$) corresponds to the spike train produced by a perfect
integrate-and-fire neuron with an exponential time-dependent
threshold \cite{lindner2005}.

\section{Comparison to numerical results}
\indent In this section we test different theoretical results and
compare them with numerical simulations. As shown, the expansion
given by Eq.~(\ref{eq6}) is the exact solution of the system.
However, the explicit computation of successive terms in the
expansion is performed up to certain order. Since truncation
introduces an error for any finite order, we mainly focus on the
first-order expansion with small values of $\epsilon$. In this
case, the time-dependent exponential drift can be thought of as a
perturbation to the unperturbed system defined by $\epsilon = 0$
(constant drift case). Without mathematical loss, we set all
quantities of the system to nondimensional units.

\subsection{Linear order expansion}
\begin{figure}[t!]
\begin{center}
\includegraphics[scale=0.2]{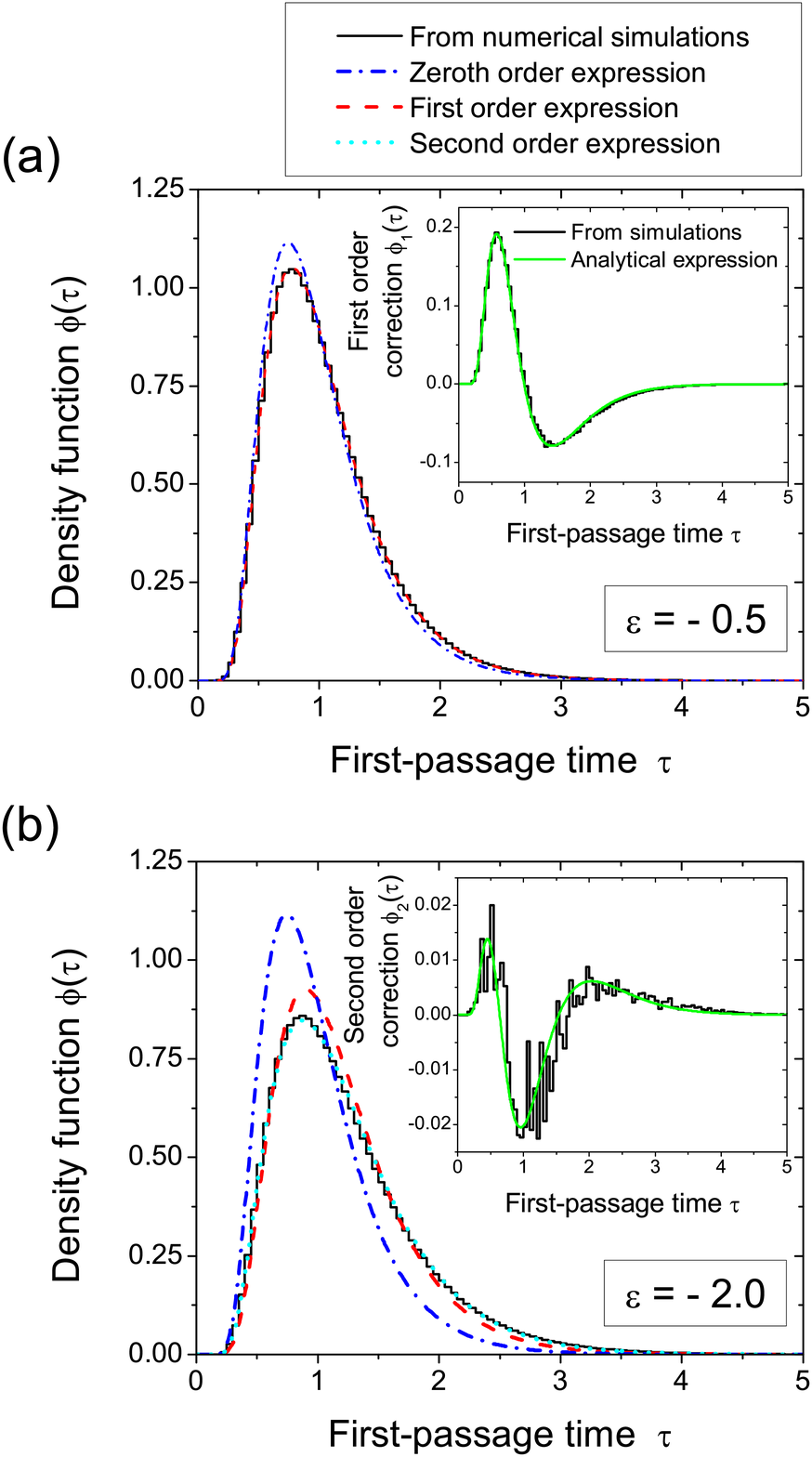}
\caption{\label{fig1} (Color online) First-passage time density
functions $\phi(\tau)$ for different intensities of the
exponential time-dependent drift. Histograms obtained from
simulations (stair-like solid lines) are compared with analytical
results: zeroth-order [dotted-dashed (blue) lines],
$\phi_{0}(\tau)$; first-order [dashed (red) lines],
$\phi_{0}(\tau) + \epsilon \phi_{1}(\tau)$; and second-order
[dotted (cyan) lines], $\phi_{0}(\tau)+\epsilon
\phi_{1}(\tau)+\epsilon^2 \phi_{2}(\tau)$, expressions.
$\phi_{0}(\tau)$ is given by Eq.~(\ref{eq42}), whereas
$\phi_{1}(\tau)$ and $\phi_{2}(\tau)$ are obtained from numerical
Laplace inversion of the corresponding expressions
[Eq.~(\ref{eq34}) for $\phi_{1}(\tau)$ and Eqs.~(\ref{ad5}) and
(\ref{eq32}) for $\phi_{2}(\tau)$]. (a) Low intensity, $\epsilon =
-0.5$; (b) high intensity, $\epsilon = -2.0$. Insets: The
analytical expression [solid (green) line] for the first-(second-)
order function $\phi_{1}(\tau)$ [$\phi_{2}(\tau)$] is compared
with the empirical linear (second-order) function (stair-like
solid lines), see text for definition. Remaining parameters are $N
= 10^7$ simulations for each case, $\mu = 1.0$, $D = 0.1$,
$x_{\text{thr}}-x_0 = 1.0$, and $\tau_{\text{d}} = 10.0$.}
\end{center}
\end{figure}

\indent In Figs.~\ref{fig1}(a) and \ref{fig1}(b) we show the FPT
density obtained from simulations for different intensities of the
exponential drift [(a) $\epsilon = - 0.5$ and (b) $\epsilon =
-2.0$]. The histogram obtained from simulations (stair-like solid
line) is compared with different predictions. The zeroth-order
prediction, $\phi_{0}(\tau)$, is given by the inverse Laplace
transform of Eq.~(\ref{eq33})

\begin{equation}\label{eq42}
   \phi_{0}(\tau) = \frac{(x_{\text{thr}}-x_0)}{\sqrt{4\pi D \tau^3}}
   \exp\Big\{ -\frac{\left[(x_{\text{thr}}-x_0)-\mu
   \tau\right]^2}{4D\tau}\Big\},
\end{equation}

\noindent and corresponds to the constant drift case ($\epsilon =
0$). The linear-order solution is composed of $\phi_{0}(\tau) +
\epsilon \phi_{1}(\tau)$. The function $\phi_{1}(\tau)$ is
obtained from numerical Laplace inversion of Eq.~(\ref{eq34}).\\
\indent As shown in Fig.~\ref{fig1}(a), for low intensities of the
exponential drift, the FPT statistics is well characterized by the
linear order. As expected, when the intensity is increased,
higher-order effects become significant and the linear expansion
is not enough [Fig.~\ref{fig1}(b)]. In this case, the second-order
solution, given by Eqs.~(\ref{ad5}) and (\ref{eq32}) and numerical
inverse Laplace transformation, successfully accounts for the
numerical data.

\indent To directly compare the linear correction $\phi_{1}(\tau)$
with its numerical equivalent, we construct an empirical linear
function as follows. From $N$ FPT processes for the constant
($\epsilon = 0$) and time-dependent ($\epsilon \neq 0$) cases, we
obtained their histograms, $\phi_{\text{const}}(\tau)$ and
$\phi_{\text{timedep.}}(\tau)$, and define
$\phi_{1}^{\text{empir}}(\tau) = [\phi_{\text{timedep.}}(\tau) -
\phi_{\text{const}}(\tau)] / \epsilon$. Obviously,
$\phi_{\text{const}}(\tau)$ coincides with $\phi_{0}(\tau)$ (not
shown). If the FPT process is dominated by the linear regime, the
empirical function so obtained should agree with the analytical
result, $\phi_{1}(\tau)$. As shown in the inset in
Fig.~\ref{fig1}(a), both functions coincide for a small
perturbation $\epsilon = -0.5$ [stair-like line represents the
empirical function, whereas the solid (green) line is the
analytical expression] and are a relative mismatch for $\epsilon =
-2.0$ (not shown). In this case, given the empirical linear
correction constructed with a tiny perturbation $\epsilon = -0.1$,
we can construct a similar empirical second-order correction
function as $\phi_{2}^{\text{empir}}(\tau) =
[\phi_{\text{timedep.}}(\tau) - \phi_{\text{const}}(\tau) -
\epsilon~\phi_{1}^{\text{empir}}(\tau)] / \epsilon^{2}$. As shown
in the inset in Fig.~\ref{fig1}(b) this empirical function
coincides with its analytic counterpart $\phi_{2}(\tau)$
(fluctuations due to a finite number of simulations become higher
than in the linear construction).\\
\indent Given the parameters $\mu = 1$ and $(x_{\text{thr}}-x_0) =
1.0$, the mean FPT for the unperturbed case is $1$. As shown in
the inset in Fig.~\ref{fig1}(a), a small additive exponential
drift generates a biphasic correction to the constant case
density, with a positive weight for times shorter than the
unperturbed mean and a negative weight for longer times. The
overall effect is to distort the density, increasing the
probability of times shorter than the typical time in the
unperturbed system leaving the survival domain, for positive
perturbations. For negative perturbations, the contrary is true.

\subsection{Moments of the first-passage time density}
\begin{figure}[t!]
\begin{center}
\includegraphics[scale=0.212]{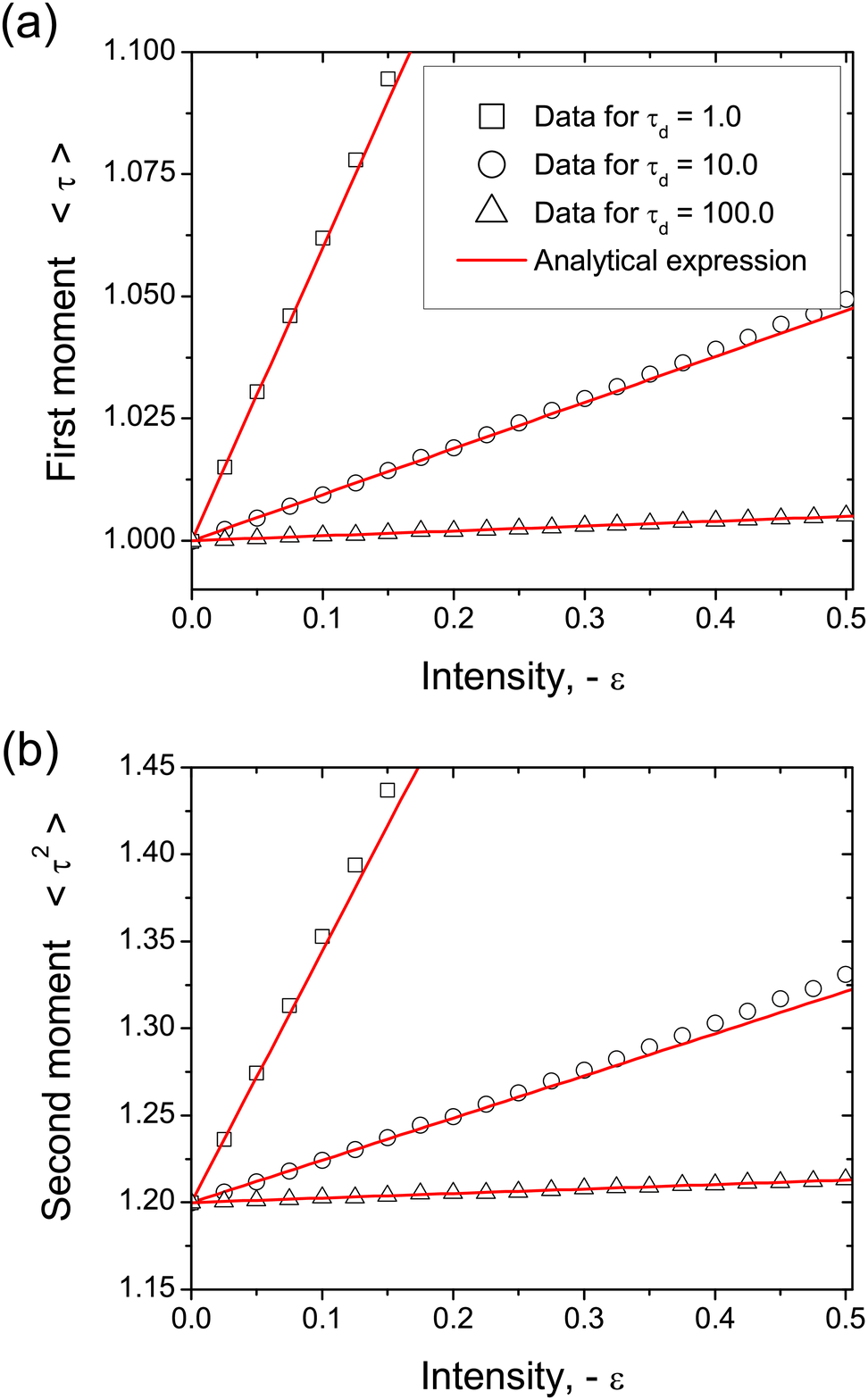}
\caption{\label{fig2} (Color online) The first two moments as a
function of the (negative) intensity of the exponential
perturbation, $-\epsilon$, for different $\tau_{\text{d}}$.
Analytical expressions [solid (red) line] are given by
Eqs.~(\ref{eq38}) and (\ref{eq39}), whereas data obtained from
simulations are represented by different symbols. Remaining
parameters as in Fig.~\ref{fig1}.}
\end{center}
\end{figure}

\indent As expressed by Eq.~(\ref{eq36}), a linear correction to
the FPT density of the unperturbed system is reproduced in all its
moments. In Fig.~\ref{fig2} we show the first two moments for the
unperturbed case ($\epsilon = 0$) and different strengths of the
perturbation, $\epsilon \neq 0$. The analytical expressions for
these moments are given by $\langle \tau^{k} \rangle = \langle
\tau^{k} \rangle_{\phi_0} + \epsilon \langle \tau^{k}
\rangle_{\phi_1}$ ($k=1,2$), where $\langle \tau^{k}
\rangle_{\phi_i}$ are given by Eqs.~(\ref{eq38}) and (\ref{eq39}).
Arbitrarily, we use negative values for $\epsilon$. In this case,
moments shift toward larger values in comparison to the
unperturbed case. As shown, both moments [Figs.~\ref{fig2}(a) and
\ref{fig2}(b)] coincide with the \textit{linear} analytical
results for low intensities and mismatch for larger values. To
describe these properties properly for large values of $\epsilon$,
we should include higher-order terms in the expansion given,
Eq.~(\ref{eq36}). The range of validity of the linear regime is
given by the time scale of the exponential perturbation
$\tau_{\text{d}}$, which set the effective intensity in the
Langevin equation [see Eq.~(\ref{eq1})]. As the time scale of the
exponential drift increases, the linear regime remains valid over
a wider range in the $\epsilon$ coordinate.

\begin{figure}[t!]
\begin{center}
\includegraphics[scale=0.212]{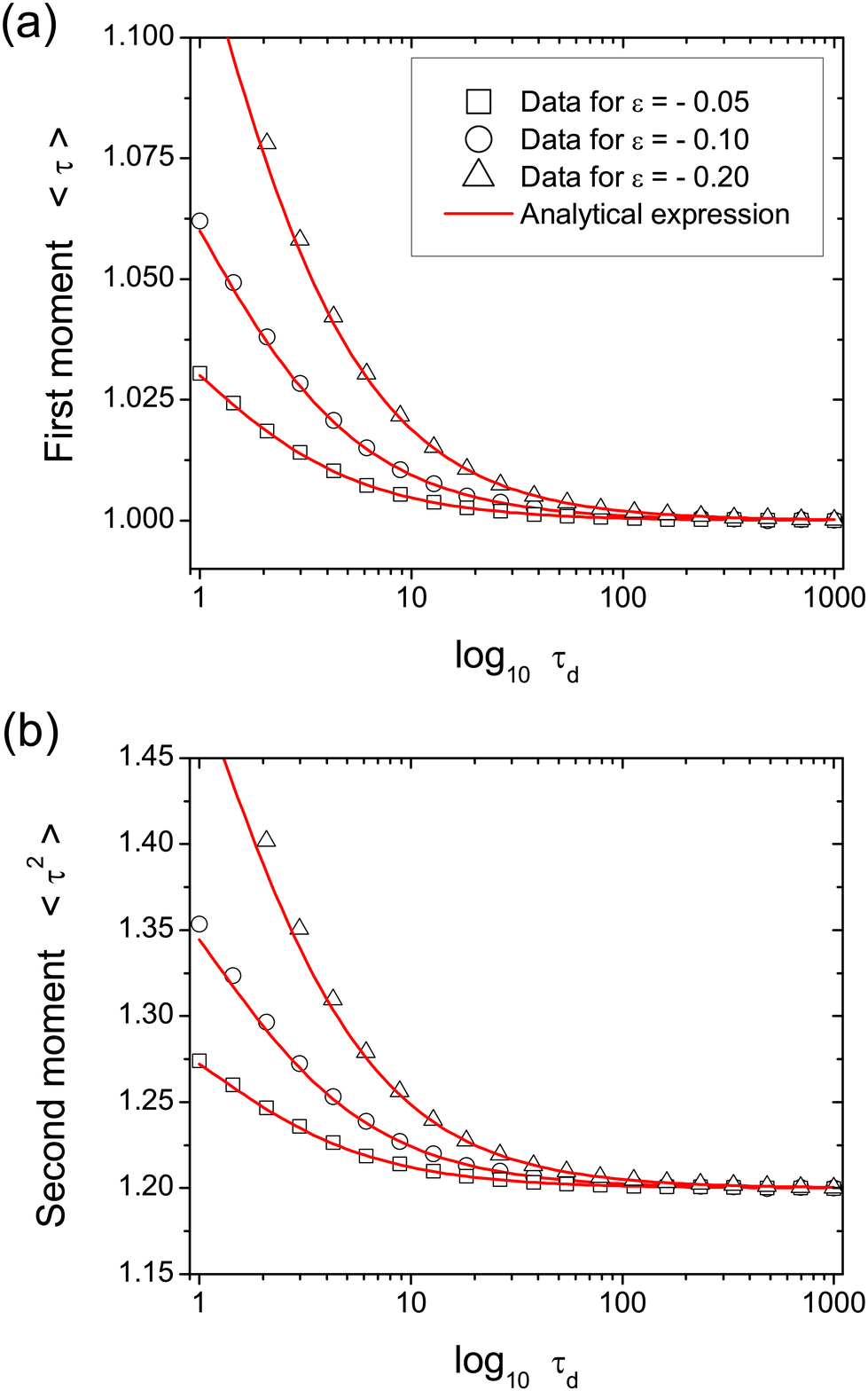}
\caption{\label{fig3} (Color online) The first two moments as a
function of the exponential time scale $\tau_{\text{d}}$
(semi-logarithmic plot), for different strengths of the
perturbation. Remaining data as in Fig.~\ref{fig2}.}
\end{center}
\end{figure}

\indent To stress the preceding paragraph, in Fig.~\ref{fig3} we
show the first two moments as a function of the time scale
(logarithmic scale) for different perturbation intensities. In
this case, a given perturbation $\epsilon$ produces a linear
distortion for a large time scale $\tau_{\text{d}}$, but
higher-order effects become important for smaller time constants.
However, as we see in the last subsection, the results we obtained
for the linear expansion still hold in the limit of vanishing time
scales $\tau_{\text{d}} \rightarrow 0$.

\subsection{Spectral density}
\begin{figure}[t!]
\begin{center}
\includegraphics[scale=0.212]{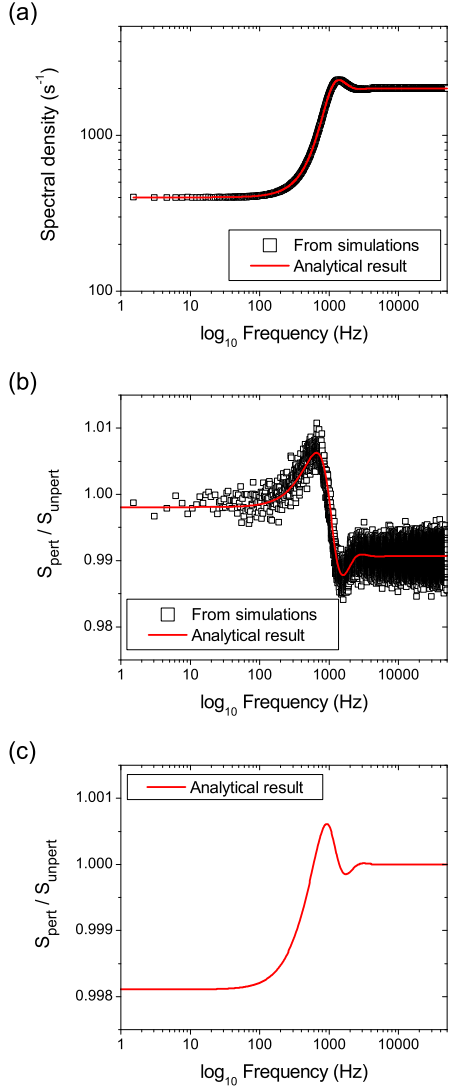}
\caption{\label{fig4} (Color online) (One-sided) Spectral density
(per unitary time) of spike trains for different cases. In each
case, the spectral density was obtained as an average of $10^6$
independent simulations. (a) Spectral density of the unperturbed
case ($\epsilon = 0$). Parameters: $\mu = 1.0~\text{ms}^{-1}$, $D
= 0.1~\text{ms}^{-1}$, $x_{\text{thr}}-x_0 = 1.0$. (b) Ratio
between spectral densities, $S_{\text{pert}}(\omega) /
S_{\text{unpert}}(\omega)$, for the exponential time-dependent
drift as a perturbation. Same parameters as in (a) for each case,
and the perturbed system is additionally defined by $\epsilon =
-0.1$ and $\tau_{\text{d}} = 10.0~\text{ms}$. (c) Ratio between
spectral densities, $S_{\text{pert,mc}}(\omega) /
S_{\text{unpert}}(\omega)$, for the mean corrected exponential
drift (see text). Same parameters as in (a) for the unperturbed
case. For the exponential drift case, $D$ and $x_{\text{thr}}-x_0$
are the same as in the unperturbed case, and the exponential drift
is characterized by $\epsilon = -0.1$ and $\tau_{\text{d}} =
10.0~\text{ms}$ [same as in (b)], but the constant component of
the drift, $\mu_{\text{pert,mc}}$ is changed in order to obtain
the same mean $\langle \tau \rangle$ as in the unperturbed case
($\langle \tau \rangle = 1.0~\text{ms}$).}
\end{center}
\end{figure}

\indent To simplify the following discussion, here we set time
units in milliseconds, while maintaining $x$ as a nondimensional
magnitude. In this case, $\mu$, $D$, and $\tau_{\text{d}}$ are
measured as $1/$milliseconds, $1/$milliseconds, and milliseconds,
respectively. As indicated in Sec. \ref{prop}, the spectral
density of a renewal process composed of consecutive first
passages (hereafter called the spike train) is easily computed
with the Laplace transform of the density function,
Eq.~(\ref{eq41}). For example, the (one-sided) spectral density
(per unitary time) \cite{num_recipes} of the unperturbed system
has a relatively simple expression (see Eq.~(3.17) in
\cite{stein1972}), which is shown in Fig.~\ref{fig4}(a).

\indent As indicated at the beginning of this section, we restrict
ourselves to consideration of the effect of a \textit{small}
additive exponential time-dependent perturbation on the spectral
properties of the spike train evoked by a system driven by a
leading constant drift $\mu$. For such a situation, the change in
the spectral density of the unperturbed system is hardly
noticeable, and therefore, to analyze the frequency-dependent
changes introduced by the perturbation, we focus on the ratio
between spectral densities. If the unperturbed system is
characterized by $\mu$, $D$ and $x_{\text{thr}}-x_0$
[Fig.\ref{fig4}(a)], and an exponential drift is added to the
system, defined by $\epsilon$ and $\tau_{\text{d}}$, the effect of
this perturbation on the spectral density is shown in
Fig.~\ref{fig4}(b). In this case, we set a negative perturbation
$\epsilon < 0$, which implies that the mean FPT increases [see
Eq.~(\ref{eq39})]. Consequently, the rate of the consecutive
first-passage processes decreases to lower frequencies, in
comparison with the unperturbed case. This rate is strictly given
by the value of the spectral density at infinite (two distant
events are uncorrelated, which means a ``white'' spectrum at
frequencies tending to infinity, for stationary processes),
$S(\omega \rightarrow \infty)$, and it roughly determines the
position of the observed peak [see Fig.~\ref{fig4}.(a)]. For
example, in Fig.~\ref{fig4}(a) the rate is defined by $\langle
\tau \rangle^{-1} = \mu/(x_{\text{thr}}-x_0) = 1~\text{ms}^{-1}$.
This results in an asymptotic value of $10^3~\text{s}^{-1}$ [a
factor equal to $2$ appears when considering the
\textit{one-sided} spectral density \cite{num_recipes}, such as
that shown in Fig.~\ref{fig4}(a)] and a peak located near
$10^3~\text{Hz}$. The lower rate obtained by the addition of a
negative exponential perturbation decreases the asymptotic
spectral value and shifts the peak to a lower frequency. In the
spectral ratio we consider in Fig.~\ref{fig4}(b),
$S_{\text{pert}}(\omega)/S_{\text{unpert}}(\omega)$, these effects
are reflected by an asymptotic value less than unity and a
biphasic shape, with a positive (negative) peak located at a lower
(higher) rate than the unperturbed rate.\\
\indent The effects already mentioned (representative of the
linear regime) are mainly related to the mean $\langle \tau
\rangle$, so we consider an alternative situation where this
property does not change. In this case, if the unperturbed system
has a given mean [in Fig.~\ref{fig4}(a), $\langle \tau \rangle =
1~\text{ms}$], the perturbed system will be driven by the low
exponential time-dependent drift, freely defined by $\epsilon$ and
$\tau_{\text{d}}$, and the constant component will be modified (in
comparison to the unperturbed system) in order to keep the mean
unchanged. In consequence, the asymptotic value for the ratio
between the spectra should be equal to unity, as shown in
Fig.~\ref{fig4}(c). In this figure we note that other effects are
present since the ratio is not flat, but they are an order of
magnitude less than the case where the exponential term is a
direct additive effect to the unperturbed system
[Fig.~\ref{fig4}.(b)]. Moreover, these effects are hardly
noticeable by simulations (even for the large set we used).
Therefore, the change in the mean is the main effect introduced by
this kind of time-dependent perturbation.

\subsection{Limit behaviors}
\indent Finally, here we analyze the behavior of the first-order
analytical solution we explicitly obtained, for large and small
time scales of the exponential drift.

\subsubsection{Limit $\tau_{\text{d}} \rightarrow \infty$}
\indent In section \ref{theory} we obtained the FPT density
function as an expansion in $\epsilon$ and explicitly derived the
first-order expression [for the FTP density, it is easy to obtain
the second-order expression from Eq.~(\ref{ad5}) via
Eq.~(\ref{eq32})]. In particular, we have this solution fully
characterized in the Laplace domain. To analyze the limit of large
time scales, $\tau_{\text{d}} \rightarrow \infty$, we expand
Eq.~(\ref{eq34}) in terms of $(1/\tau_{\text{d}})$ and keep the
lowest terms. Up to order $1$, this expansion reads

\begin{eqnarray}\label{eq43}
   \tilde{\phi}_{1}^{L}(s) &\approx&
   -\frac{(x_{\text{thr}}-x_0)}{2D\tau_{\text{d}}}~\frac{\mu-\sqrt{\mu^2 + 4Ds}}{\sqrt{\mu^2 +
   4Ds}}\nonumber\\
   &&\cdot \exp\Big\{ \frac{(x_{\text{thr}}-x_0)}{2D}\left[ \mu-\sqrt{\mu^2+4Ds}
   \right]\Big\}.
\end{eqnarray}

\indent From this expression we can obtain, by differentiation in
$s$ and evaluation at $s=0$, the linear term in the expansion of
the moments of the density function, valid for this limit. The
resulting expressions coincide with those in
\cite{lindner2004}.\\
\indent Eq.~(\ref{eq43}) shows the Laplace transform of the linear
term appearing in the $\epsilon$ expansion, for the limit
$\tau_{\text{d}} \rightarrow \infty$. In particular, it is
possible to find its inverse Laplace transform
\cite{oberhettinger}, which reads

\begin{eqnarray}\label{eq44}
   \phi_{1}^{(\tau_{\text{d}}\rightarrow \infty)}(\tau) &=&
   \frac{(x_{\text{thr}}-x_0)}{\sqrt{4\pi D \tau^3}} ~ \frac{(x_{\text{thr}}-x_0)-\mu \tau}{2D
   \tau_{\text{d}}}\nonumber\\
   &&\cdot \exp\Big\{ -\frac{\left[(x_{\text{thr}}-x_0)-\mu
   \tau\right]^2}{4D\tau}\Big\}\nonumber\\
   &=& \frac{(x_{\text{thr}}-x_0)-\mu \tau}{2D
   \tau_{\text{d}}} \cdot \phi_{0}(\tau).
\end{eqnarray}

\indent In Fig.~\ref{fig5}(a) we show the product of the
asymptotic linear correction function, Eq.~(\ref{eq44}),
multiplied by $\tau_{\text{d}}$ as a function of $\tau$. This
product does not depend on $\tau_{\text{d}}$ and, therefore, can
be compared on the same scale with the products obtained from
simulations. As expected, as $\tau_{\text{d}}$ increases, both the
empirical product and the analytical result coincide.\\
\indent From Eq.~(\ref{eq44}), it is easy to see that the FPT
density function in this limit, up to order $1$, is

\begin{equation}\label{eq45}
   \phi(\tau) = \left[ 1 + \epsilon ~\frac{(x_{\text{thr}}-x_0)-\mu\tau}{2D\tau_{\text{d}}}
   \right]~ \phi_{0}(\tau).
\end{equation}

\indent In fact, instead of working out the limit $\tau_{\text{d}}
\rightarrow \infty$ in Eq.~(\ref{eq34}) as we did before, we can
look at the actual physical situation. For $\tau_{\text{d}}
\rightarrow \infty$, the exponential drift can be thought of as a
constant (i.e., $\tau_{\text{d}} \gg \langle \tau \rangle$). In
this case, the constant drift would be $\mu + \epsilon /
\tau_{\text{d}}$ and the FPT density function would be given by
$\phi_{0}(\tau)$, Eq.~(\ref{eq42}), with this modified drift:

\begin{equation}\label{eq46}
   \phi(\tau) = \frac{(x_{\text{thr}}-x_0)}{\sqrt{4\pi D\tau^3}}~\exp\Big\{ -\frac{\left[(x_{\text{thr}}-x_0)-(\mu
   +\epsilon/\tau_{\text{d}}) \tau \right]^2}{4D\tau}\Big\}.
\end{equation}

\indent Expanding Eq.~(\ref{eq46}) around $\epsilon = 0$ up to
order $1$ gives the same result as obtained before,
Eq.~(\ref{eq45}).

\subsubsection{Limit $\tau_{\text{d}} \rightarrow 0$}
\indent In this limit, the first exponential term between the
large curly brackets in Eq.~(\ref{eq34}), which contains the
expression $(s+1/\tau_{\text{d}})$, vanishes. The reason is that
the real part of the exponent tends to $-\infty$ as
$\tau_{\text{d}}$ tends to $0$. Therefore, the linear correction
term of the density function, in the Laplace domain, simplifies to

\begin{eqnarray}\label{eq47}
   \tilde{\phi}_{1}^{L}(s) &=& - \frac{\mu-\sqrt{\mu^2+4Ds}}{2D} \nonumber\\
   &&\exp\Big\{ \frac{(x_{\text{thr}}-x_0)}{2D}\left[ \mu-\sqrt{\mu^2+4Ds}
   \right]\Big\}.
\end{eqnarray}

\indent As in the previous limit, the linear term in the expansion
of the moments can be obtained directly from Eq.~(\ref{eq47}) and
coincide with the corresponding case in \cite{lindner2004}.\\
\indent The simplified expression obtained, Eq.~(\ref{eq47}), is
analytically tractable and the inverse Laplace transform is easy
to compute \cite{oberhettinger}. In the temporal domain, the
linear correction is

\begin{eqnarray}\label{eq48}
   \phi_{1}^{(\tau_{\text{d}}\rightarrow 0)}(\tau) = \frac{(x_{\text{thr}}-x_0)}{\sqrt{4\pi
   D\tau^3}}~\exp\Big\{ -\frac{\left[(x_{\text{thr}}-x_0)-\mu
   \tau\right]^2}{4D\tau}\Big\}\nonumber\\
   \cdot \frac{1}{2D\tau}\left[ (x_{\text{thr}}-x_0) -\mu\tau -
   \frac{2D\tau}{(x_{\text{thr}}-x_0)}\right]\hspace{0.3cm}
   \nonumber\\
   = \frac{\phi_{0}(\tau)}{2D\tau}\left[ (x_{\text{thr}}-x_0) -\mu\tau -
   \frac{2D\tau}{(x_{\text{thr}}-x_0)}\right]. \hspace{0.175cm}
\end{eqnarray}

\indent In Fig.~{\ref{fig5}(b) we can observe this limit
expression as a function of $\tau$. Additionally, simulations
based on different $\tau_{\text{d}}$ values show that the limit is
reached for sufficiently small values.

\begin{figure}[t!]
\begin{center}
\includegraphics[scale=0.2]{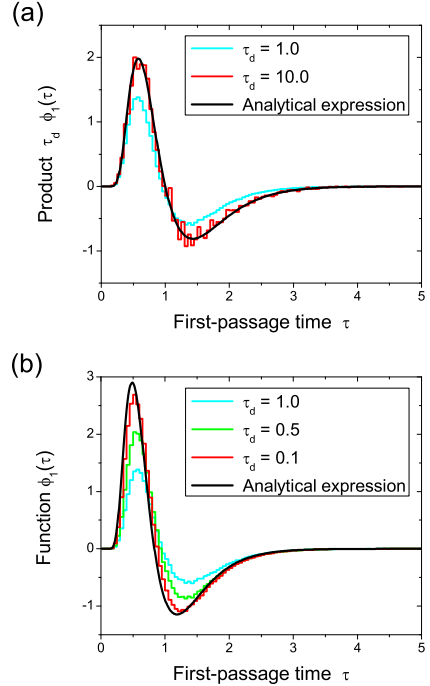}
\caption{\label{fig5} (Color online) Limit behaviors for the
linear correction function $\phi_{1}(\tau)$. (a) In the limit
$\tau_{\text{d}} \rightarrow \infty$ the product $\tau_{\text{d}}
\cdot \phi_{1}(\tau)$ does not depend on $\tau_{\text{d}}$ [solid
(black) line], Eq.~(\ref{eq44}). The empirical product is shown
for different $\tau_{\text{d}}$ values (colored stair-like lines).
As $\tau_{\text{d}}$ increases, the empirical histogram (obtained
as in the insets in Fig.~\ref{fig1} multiplied by the
corresponding $\tau_{\text{d}}$) coincides with the analytical
expression. (b) In the limit $\tau_{\text{d}} \rightarrow 0$ the
density function $\phi_{1}(\tau)$ does not depend on
$\tau_{\text{d}}$ [solid (black) line], Eq.~(\ref{eq48}). The
empirical function is shown for different $\tau_{\text{d}}$ values
(colored stair-like lines). As $\tau_{\text{d}}$ decreases, the
empirical histogram tends to the analytical expression. Parameters
of the simulation as in Fig.~\ref{fig1}, with $\epsilon=-0.1$ and
different $\tau_{\text{d}}$ values.}
\end{center}
\end{figure}

\indent From Eq.~{\ref{eq48}}, we can compute the FPT density
function for this limit which, up to order $1$, reads

\begin{equation}\label{eq49}
   \phi(\tau) = \Bigg\{ 1 + \frac{\epsilon}{2D\tau} \left[ (x_{\text{thr}}-x_0) -\mu\tau -
   \frac{2D\tau}{(x_{\text{thr}}-x_0)} \right] \Bigg\}
   ~\phi_{0}(\tau).
\end{equation}

\indent The limit $\tau_{\text{d}} \rightarrow 0$ also enables a
physical interpretation. Since $\exp[-(t-t_0) / \tau_{\text{d}}] /
\tau_{\text{d}} \rightarrow \delta(t_0)$ as $\tau_{\text{d}}
\rightarrow 0$, from Eq.~(\ref{eq1}) it is easy to see that after
a differential time from the initial time $t_0 + dt$, the system
has moved to the position $x_{0}+\epsilon$, and thereafter, the
dynamics is governed by a constant drift $\mu$. In this case, the
FPT density function is $\phi_{0}(\tau)$, Eq.~(\ref{eq42}), with
the initial position modified:

\begin{equation}\label{eq50}
   \phi(\tau) = \frac{(x_{\text{thr}}-x_0-\epsilon)}{\sqrt{4\pi
   D\tau^3}}~\exp\Big\{ -\frac{\left[(x_{\text{thr}}-x_0-\epsilon)-\mu
   \tau\right]^2}{4D\tau}\Big\}.
\end{equation}

\indent The expansion of Eq.~(\ref{eq50}) around $\epsilon=0$ up
to order $1$ also coincides with Eq.~(\ref{eq49}).\\

\section{Discussion and concluding remarks}
\indent In the present work we have analyzed the survival
probability and the FPT problem of a Wiener process driven by an
exponential time-dependent term superimposed to a constant drift,
Eq.~(\ref{eq1}), in the presence of an absorbing fixed boundary.
We first focus on the survival probability in the region of
interest and derive the time-inhomogeneous diffusion equation
governing it, Eq.~(\ref{eq5}), in the framework of the backward FP
formalism. We propose a solution given by an expansion in terms of
the intensity of the exponential drift, Eq.~(\ref{eq6}), and
derive the associated equations and (boundary and initial)
conditions to solve each term, Eqs.~(\ref{eq8}) to (\ref{eq11}).
Interestingly, the resulting equations are recurrent and easy to
solve via a Laplace transformation. We explicitly solve up to the
second-order term in the expansion, in the Laplace domain, and
give some remarks about higher-order terms (see corresponding
subsections). In particular, we show that each term exists, and
therefore, the expansion we proposed is justified and constitutes
the exact solution. Moreover, when the solution is set to the
initial conditions of the problem, the probability depends
exclusively on the time elapsed from the initial time, as expected
from physical considerations, Eq.~(\ref{eq23}). The FPT density
function is obtained in terms of the survival probability, and we
show that the expansion is preserved in this function and its
moments, Eqs.~(\ref{eq28}) and (\ref{eq36}). Since the solution of
each term is easily obtained in Laplace domain and the inverse
transform is not always available, we review some related
properties that can be calculated from them: the moments of the
density function and the spectral density of an associated renewal
process or ``spike train''.\\
\indent In the second part of this work, we focus on the
comparison between the explicit results we have obtained and
numerical simulations. Since truncation of the series results in
an approximate solution, we mainly focus on the first-order
expansion. This linear regime coincides with a perturbation
scenario. As shown in Fig.~\ref{fig1}(a), the first-order term in
the expansion of the FPT density completely defines a slightly
perturbed system, whereas a higher intensity of the exponential
time-dependent drift facilitates higher-order effects
[Fig.~\ref{fig1}(b)]. Linear expansion of the first two moments of
the density function reproduces numerical results accurately,
except for low time constants of the exponential drift
(Figs.~\ref{fig2} and \ref{fig3}). For a small exponential drift,
we calculate the spectral density of the resulting spike train and
observe that analytical results coincide extremely well
[Fig.~\ref{fig4}(b)]. Moreover, the change in the spectral
properties due to a small perturbation can be mostly ascribed to
the change in the mean FPT [Fig.~\ref{fig4}(c)]. Finally, we
derive the behavior of the linear expansion of the FPT density
function, in the limit of negligible as well as extremely large
time scales, for the time constant of the exponential drift. These
limit expressions are inverse Laplace transformed and compared
with exact results obtained from physical considerations,
valid strictly in the corresponding limit (Fig.~\ref{fig5}).\\
\indent The process considered in this work naturally arises in
neuroscience, but it is not restricted to this field (e.g.,
consider the motion of a charged particle in an exponentially
decaying electrical field). In the context of neuroscience, the
behavior of stochastic spiking neurons with an adaptation current
can be described by stochastic processes with an exponentially
decaying temporal term \cite{benda2010}. For spiking neurons, the
state variable $x$ corresponds to the membrane potential and its
evolution is given by a Langevin equation, where the integration
of an input current is performed until a threshold is reached. At
this moment a spike is generated and the time corresponds to the
FPT in the statistical description \cite{gerstner, burkitt}. The
Wiener process is the stochastic representation of the basic model
in theoretical studies, namely, the perfect integrate-and-fire
neuron. In addition to external signals, an adapting neuron
integrates, between spikes, an exponential time-dependent current
corresponding to specific ionic channels \cite{adapt}. Therefore,
the statistical description of the model we have considered
provides important measures for analyzing further effects of
adaptation on spiking neurons.\\
\indent Finally, we note some features about the methodology
considered here. First, we derive the equation governing the
transition probability for a specific temporal drift,
Eq.~(\ref{eq5}), and propose a solution in terms of a certain
expansion, Eq.~(\ref{eq6}). Both procedures can be used for any
time-dependent drift. However, in general, the resulting equations
would have a source term difficult to tackle analytically. Second,
our starting model is the Wiener process with an exponentially
decaying temporal drift. It is easy to check that an equivalent
formulation can be performed to other one-dimensional processes
(e.g., Ornstein-Uhlenbeck process) with the same kind of temporal
drifts. In these cases, the only change with respect to the Wiener
process is that the homogeneous parts of the differential
equations we have obtained, on the left-hand side of
Eqs.~(\ref{eq8}) and (\ref{eq9}), are different. Obviously, the
difficulty in solving these other cases depends on the system at
hand.

\section{Acknowledgments}
The author thanks Adrian Budini for a critical reading of the
manuscript and useful comments. This work was supported by the
Consejo de Investigaciones Cient\'ificas y T\'ecnicas de la
Rep\'ublica Argentina.

\end{document}